\documentclass[pra,aps,showpacs]{revtex4}
\usepackage{epsfig}

\begin{document}

\title{Treatment of backscattering in a gas of interacting fermions confined to
a one-dimensional harmonic atom trap}

\date{\today}

\author{Gao Xianlong, F. Gleisberg, F. Lochmann, and W. Wonneberger}

\affiliation{Abteilung f\"ur Mathematische Physik, Universit\"at Ulm, D89069 Ulm, Germany}

\begin{abstract}
An asymptotically exact many body theory for spin polarized interacting
fermions in a one-dimensional harmonic atom trap is developed using the bosonization
method and including backward scattering. In contrast to the Luttinger model,
backscattering in the trap generates one-particle potentials which must be
diagonalized simultaneously with the two-body interactions.
Inclusion of backscattering becomes necessary because backscattering is the dominant
interaction process between confined identical one-dimensional fermions. The bosonization
method is applied to the calculation of one-particle matrix elements at zero temperature.
A detailed discussion of the validity of the results from bosonization is given, including
a comparison with direct numerical diagonalization in fermionic Hilbert space.
A model for the interaction coefficients is
developed along the lines of the Luttinger model with only one coupling constant $K$.
With these results, particle densities, the Wigner function, and the central pair
correlation function are calculated and displayed for large fermion numbers. It is shown
how interactions modify these quantities. The anomalous dimension of the pair correlation
function in the center of the trap is also discussed and found to be in accord with the
Luttinger model.

\end{abstract}


\maketitle

\section{Introduction}

The achievement of Bose-Einstein condensation in dilute ultracold gases \cite{ADB95}
stimulated the theoretical interest in trapped fermionic many body systems
\cite{BDL98,BB98,ZGOM00},
especially their superfluid properties \cite{HFSt97,BP98,HSt99,C99}.
Recent experimental successes in obtaining degeneracy in three-dimensional Fermi
vapors \cite{DMJ99,OHara00,SFCCKMS01,Tru01,SK01} intensified the interest in confined Fermi gases.

Using microtrap technology \cite{VFP98,FGZ98,DCS99,RHH99,Ott01}, it is conceivable
to realize a neutral ultracold quantum gas of trapped quasi one-dimensional degenerate
fermions.

In many cases, identical spin polarized fermions experience only a
weak residual interaction because s-wave scattering is forbidden.
This makes the question of interactions in a one-component spin
polarized fermion system somewhat academic. Possible exceptions
are Feshbach resonance enhanced scattering between atoms
\cite{R02} and electric dipole--dipole interactions in the case of
polar molecules \cite{BBC00}.

The confinement of a trapped ultracold gas can be realized by a harmonic potential.
We have developed an asymptotically exact theory of interacting one-dimensional fermions
confined to a harmonic trap. It is based on the bosonization method known from the theory
of Luttinger liquids (cf. \cite{E79,Haldane,V95,Sch95}) and exploits the linearity of the
energy spectrum of free oscillator states. The method was presented in \cite{WW01} for the
one-component gas with forward scattering and extended in \cite{GaW02} to the case of
two components. This model must be seen as a soft boundary alternative to the well studied
case of one-dimensional interacting fermions confined by hard walls
\cite{C84,EA92,FG95,EG95,WVF96,MEJ97,VYG00}.

Meanwhile, an investigation of the interaction matrix elements for one-dimensional
identical fermions in the harmonic trap revealed \cite{GW02} that backscattering dominates
forward scattering unless the pair potential is long ranged. Unlike in the case
of the Luttinger model, backscattering cannot be taken into account by merely renormalizing
the forward coupling constants.
This is due to the one-branch structure of the present model, which generates one-particle
potentials from the backward scattering when the latter is brought into the form of an
effective forward scattering. These one-particle potentials must be diagonalized simultaneously
with the two particle interactions from forward scattering. We solve this problem by
supplementing the squeezing transformation with an appropriate displacement
transformation.

The paper is organized as follows: Section II describes the model and classifies the scattering
processes for fermions in the one-dimensional harmonic trap. Section III gives the solution of
the backscattering problem for the one-particle matrix elements using the bosonization method.
Section IV discusses the validity of the bosonization scheme and compares the results with
those from direct numerical diagonalization in fermionic Hilbert space.
Sec. V presents results for several quantities of interest: Occupation probabilities,
particle and momentum densities, and the central pair correlation function, which are
all derivable from the one-particle matrix elements. We employ a model of the interaction
coefficients developed in analogy with the Luttinger model. It is characterized by just
one coupling constant $K$ and by a small decay parameter $r \ll 1$. This approach is described
in the Appendix.

\section{Theory}

\subsection{Description of the model}

We consider spin polarized fermionic atoms interacting via the pair interaction operator

\begin{eqnarray}\label{1.0}
\hat{V} =\frac{1}{2} \sum _{mnpq}
V(m,p;q,n)\,(\hat{c}^\dagger_m \hat{c}_q ) ( \hat{c}^\dagger_p \hat{c}_n ).
\end{eqnarray}
The fermions are confined to a highly anisotropic axially symmetric harmonic trap. The trap
potential is

\begin{equation}\label{1.1}
V(x,y,z) = \frac{1}{2} m_A \omega ^2 _{\ell} \,z^2 +\frac{1}{2} m_A \omega _\perp^2 \,(x^2+y^2)
\equiv V_z(z)+V_\rho(\rho).
\end{equation}
 The atom mass is denoted $m_A$ and $z$ is the one-dimensional coordinate in the
 elongated axial direction of the trap. The trap frequencies are $\omega _\ell$ and
 $\omega _\perp
 \gg \omega _\ell$. The quasi one-dimensional Fermi gas is characterized by
 $N$ ($\le \omega _\perp /\omega _\ell$) identical fermions filling the first $N$ one-particle
 levels

 \begin{equation}\label{1.2}
 \hbar \omega _n = \hbar \omega _{\ell} (n+1/2),\,\,\, n=0,1,...
 \end{equation}
 of the one-dimensional harmonic potential $V_z$,
 while the transverse part of each unperturbed wave function remains the transverse ground
 state $\psi _{\perp 0}(x)\psi _{\perp 0}(y)$.

 The unperturbed one-dimensional Fermi energy is

 \begin{eqnarray}\label{1.3}
 \epsilon _F = \hbar \omega _\ell (N-1) + \frac{1}{2} \hbar
 \omega _\ell =  \hbar \omega _\ell (N-\frac{1}{2}).
 \end{eqnarray}
 The Fermi energy is only slightly smaller than the excitation energy $\hbar \omega _\perp$
 of the first excited transverse
 state in the case of $N=N_{max}=$ largest integer in $\omega _\perp/\omega _\ell$.
 We have in mind a situation when the filling factor $F \equiv N/N_{max}$ is small enough
 so that the Fermi level is well below this excitation energy. The above assumption about
 the transverse wave function is then justified. Furthermore, there still exists a
 macroscopic number of possible excitations of the one-dimensional Fermi sea that do
 not violate this condition.

 The operators in $\hat{c}^\dagger_m$ and  $\hat{c}_q$ in Eq. (\ref{1.0}) denote fermion creation
 and destruction operators, respectively. They obey the fermionic algebra
 $ \hat{c}_m \hat{c}_n^\dagger + \hat{c}_n^\dagger \hat{c}_m = \delta _{m,n}$.
 This ensures that each oscillator state with single particle wave function

 \begin{equation}\label{1.4}
 \psi _n(z) = \sqrt{\frac{\alpha}{2^n n!\pi^{1/2}}}
 \,e^{-\alpha^2 z^2/2 }\,H_n (\alpha z)
 \end{equation}
 and energy $\epsilon _n = \hbar \omega _n $ is at most singly occupied.
 The intrinsic length scale of the system is the oscillator length $l = \alpha ^{-1}$
 where $\alpha$ is defined by $\alpha^2=m_A \omega
 _\ell/\hbar$. $ H_n$ denotes a Hermite polynomial. The non-interacting Hamiltonian
 $\hat{H}_0 = \hbar \sum ^\infty _{n=0} \omega _n \hat{c}^+_n \hat{c}_n$
 has linear dispersion and incorporates exactly the harmonic trapping potential.

 The interaction matrix elements $V(m,p;q,n)$ in Eq. (\ref{1.0}) are calculated from effective
 one-dimensional pair potentials using the harmonic oscillator states (\ref{1.4}). Thus each individual interaction
 matrix element contains information about the harmonic trap.

 In a degenerate Fermi system, the most relevant states are those near the Fermi energy, i.e.
 $m \approx p \approx q \approx n \approx N$. This limited number of interaction matrix elements can be
 further reduced by a classification scheme based on approximate momentum
 conservation in the center of the trap and near the Fermi energy.
 This is described in the following subsection.

\subsection{Classification of coupling coefficients}

We are interested in the modification of the zero temperature Fermi sea due to the
interaction (\ref{1.0}). This problem is different from that of scattering between
excitations above the Fermi sea, which, in the present one-dimensional context, are
probably not quasi particles.

First of all, we must classify the coupling coefficients in Eq.
(\ref{1.0}). There are 24 coupling coefficients $V(m,p;q,n)$ for
any set of four distinct integers. Due to the symmetries $(m
\leftrightarrow q)$, $(n \leftrightarrow p)$, and
$([m,q]\leftrightarrow [n,p]) $ only three coupling coefficients
remain. It is shown in \cite{GW02} that among them, those with
index combinations $n=m+p-q$, $n=p+q-m$, and $n=m+q-p$ dominate,
provided the fermion number $N$ is large. It is here, where the
fermionic nature of the quantum gas becomes relevant.

We can thus write

\begin{eqnarray}\label{1.8}
V (m,p;q,n) \rightarrow V _a \, \delta _{m-q, n-p}+V _b \, \delta _{q-m, n-p}+
V _c \, \delta _{m+q, n+p}.
\end{eqnarray}
Other weakly inhomogeneous trapping potentials also give dominant
sets of $V_a, V_b$ and $V_c$ coupling constants but do not lead to
a linear dispersion required for strict bosonization. Note that
the coupling constants still depend on indices, e.g.
$V_c=V(m,p;q,n=m+q-p)$. In the present model, this is further
simplified to $V_c=V_c(|p-q|)$.

Qualitatively, Eq. (\ref{1.8}) can be understood as follows: The
single particle states $\psi _n$ well inside the trap are
superpositions of plane wave states $\exp( ik_n z)$ with
$k_n=\pm\alpha\, \sqrt{2n+1}$. For $N \gg 1$, the relevant states
are near the Fermi energy and thus $|k_n| \approx k_F =\alpha
\sqrt{2N-1}$. Here, $k_F$ denotes the Fermi wave number.

According to Eq. (\ref{1.0}), incoming states $\{n,q\}$ are transformed into $\{p,m\}$
in the collision process and the momenta of these states are (approximately, because
of the weakly
inhomogeneous trapping potential) conserved. Denoting a state with
$k_n \approx -k_F$ by $(-n)$, three distinct collision processes can be discriminated:

\begin{eqnarray}\label{1.8a}
\{n,q \} \rightarrow \{p,m\},\quad \{n,(-q)\} \rightarrow \{p,(-m)\},\quad
\{n,(-q)\} \rightarrow \{(-p),m \}.
\end{eqnarray}
Processes with strict momentum conservation dominate: This
explains the Kronecker symbols in the approximate relation
(\ref{1.8}). Momentum transfer in the first two cases is small,
thus describing forward scattering. In the last case, the momentum
transfer is about $2 k_F$, which corresponds to backward
scattering. The first two cases were considered in
\cite{WW01,GaW02}. The last one requires an extension of the
bosonization method, which is the aim of the present paper.

The couplings $V_a$, $V_b$, and $V_c$ are the analogues of the
Luttinger model couplings $g_4$, $g_2$, and $g_1$, respectively
\cite{E79,V95,Sch95}. In contrast to the Luttinger case, it was
found in \cite{GW02} that in a gas of identical fermions confined
to the harmonic trap, forward scattering is almost completely
suppressed so that backscattering is the dominant interaction
process, unless the pair interaction potential is of long range.
This is essentially a consequence of the Fermi algebra. In the
following, we will ignore $V_a$ and $V_b$ completely though a more
general treatment is possible.

In restricting the full set of interacting matrix elements to a
set of solvable interactions we define a simplified model which
cannot fully represent the initial problem.

For a number of properties such as anomalous dimensions, we can,
however, expect universality in the sense of the Luttinger liquid
phenomenology [21,23]. This is confirmed by our result for the
one-particle correlation function (Sec. V B) which shows Luttinger
liquid behavior in the center of the trap.

\section{Backscattering and Bosonization}

The treatment of forward scattering is described in
\cite{WW01,GaW02}. We give here only the extensions necessary for
the inclusion of the backscattering interaction coefficients

\begin{eqnarray}\label{1.8b}
V(m,p;q,n)= V_c (|q-p|)\, \delta _{m+q, n+p}.
\end{eqnarray}
Substituting these interaction coefficients into Eq. (\ref{1.0})
and reordering operators gives

\begin{eqnarray}\label{1.9}
\hat{V}_c = - \frac{1}{2} \sum _{mp,v \neq 0} V_c (|v|)\,( \hat{c}^\dagger_m \hat{c}_{m+v})\,
(\hat{c}^\dagger_p \hat{c}_{p+v})
+ \frac{1}{2} \sum _{m,v \neq 0} V_c (|v|) \,(\hat{c} ^\dagger _{m+v} \hat{c}_{m-v}).
\end{eqnarray}
We omitted terms which are proportional to the fermion number
operator by setting $V_c(0)=0$. The second term is a one-particle
operator, which appears due to the backscattering. The
perturbation Eq. (\ref{1.9}) is exactly solvable. Thus there is no
renormalization group flow of the coupling $V_c$.

The essential requirement of the bosonization method is the possibility to express
all operators entirely in terms of density fluctuation operators. It is still met
in the present case: Introducing the density fluctuation operators

\begin{eqnarray}\label{1.10}
 \hat{\rho}(p) \equiv \sum _{q} \hat{c}^\dagger_{q+p} \,\hat{c}_{q},
\end{eqnarray}
or, more conveniently, canonical boson operators related to them by:

\begin{eqnarray}\label{1.11}
\hat{\rho} (p) = \left \{
\begin{array}{lll}
\sqrt{|p|} & \hat{d}_{|p|}, & p< 0,\\
\sqrt{p}   & \hat{d}^\dagger_p,   & p> 0,
\end{array} \right.
\end{eqnarray}
it is found that bosonic commutation relations

\begin{eqnarray}\label{1.12}
[\hat{d}_m, \hat{d}^\dagger_n ] = \delta _{m,n},
\end{eqnarray}
are satisfied after introducing the anomalous vacuum (cf. e.g., \cite{Sch95,ML65}).

The bosonized form of the backscattering operator is

\begin{eqnarray}\label{1.13}
\hat{V}_c= - \frac{1}{2}\, \sum _{m > 0}m\,V_c (m)\,\left\{\hat{d}^2_m + \hat{d}^{\dagger 2}_m
\right \}
+\frac{1}{2} \sum _{m>0}\sqrt{2m}\, V_c (m)\,  \left [ \hat{d} _{2m} + \hat{d}^\dagger _{2m}
\right ].
\end{eqnarray}
It is seen that the two-particle interaction due to backward
scattering is of the same form as the forward scattering operator
$\hat{V}_b$ studied in \cite{WW01} except for a sign change, $V_b
\rightarrow -V_c$. This is in complete analogy to the Luttinger
case. However, the remaining one-particle operator produces
non-trivial changes.

In order to diagonalize the total Hamiltonian

\begin{eqnarray}\label{1.15}
\tilde{H} = \hbar \omega _\ell \sum _{m>0}  m \,\hat{d}^\dagger_m
\hat{d}_m + \hat{V}_c,
\end{eqnarray}
we perform two canonical transformations:

\begin{eqnarray}\label{1.17}
\hat{d}_m = \hat{S}^\dagger_2 \left \{ \hat{S} ^\dagger _1 \hat{f} _m \hat{S}_1 \right \} \hat{S}_2.
\end{eqnarray}
The first one

\begin{eqnarray}\label{1.18}
\hat{S}_1 = \exp \left[ \frac{1}{2} \sum _{m>0} \zeta _m (\hat{f}^2 _m - \hat{f} ^{\dagger 2}_m)
\right ],
\end{eqnarray}
is a kind of squeezing transformation and was used in \cite{WW01} to diagonalize the
two-particle interactions. It gives

\begin{eqnarray}\label{1.19}
\left \{ \hat{S} ^\dagger _1 \hat{f} _m \hat{S}_1 \right \} = \hat{f} _m \cosh \zeta _m
- \hat{f}^\dagger _m \sinh \zeta _m.
\end{eqnarray}
The second one

\begin{eqnarray}\label{1.20}
\hat{S}_2 = \exp \left[ \sum _{m>0} \eta _m (\hat{f}_m^\dagger - \hat{f}_m) \right],
\end{eqnarray}
is a displacement in order to get rid of the terms linear in the $\hat{d}$ and $\hat{d}^\dagger$
operators. The total result is

\begin{eqnarray}\label{1.21}
\hat{d}_m = \hat{f}_m \cosh \zeta _m - \hat{f} _m ^\dagger \sinh \zeta _m +
\eta _m \,\exp(- \zeta _m).
\end{eqnarray}

We find two diagonalization conditions: The standard one

\begin{eqnarray}\label{1.22}
\tanh 2 \zeta _m = - \frac{V_c(m)}{\hbar \omega _\ell},
\end{eqnarray}
and another one due to backscattering

\begin{eqnarray}\label{1.23}
\eta _m =\left \{
\begin{array}{ll}
- V_c (m/2)\,\exp(-\zeta _m)/(2 \sqrt{m} \,\epsilon _m), & m = 2n,\\\\
\hspace*{2cm} 0, & m=2n-1.
\end{array}
\right.
\end{eqnarray}

The final form of the Hamiltonian is

\begin{eqnarray}\label{1.24}
\tilde{H} =\tilde{H}_0+ \tilde{V}_c=\sum _{m>0} m\,\epsilon _m
\hat{f}^\dagger_m \hat{f}_m +\mbox{const.}\,.
\end{eqnarray}
The renormalized oscillator frequencies are

\begin{eqnarray}\label{1.25}
 \epsilon _m \equiv \sqrt{(\hbar \omega _\ell)^2 - V^2_c(m)},
\end{eqnarray}
and

\begin{eqnarray}\label{1.26}
\exp(-\zeta _m) \equiv \sqrt{K_m}
\end{eqnarray}
defines the dimensionless coupling constants

\begin{eqnarray}\label{1.27}
K_m \equiv \sqrt{\frac{\hbar \omega _\ell + V_c(m)} {\hbar \omega
_\ell -V_c (m)}}.
\end{eqnarray}

\subsection*{One-particle matrix elements}

We apply the theory to the evaluation of the one-particle matrix elements
$\langle \hat{c}^\dagger _{m}\,\hat{c} _{n} \rangle$. To this order, we follow
the steps in \cite{WW01} and introduce the bosonic field

\begin{eqnarray}\label{1.28}
\hat{\phi}^\dagger (v) = i \sum^\infty _{n = 1} \frac{1}{ \sqrt {n} }\, e^{-inv}
\hat{d}^\dagger _n \neq \hat{\phi}(v),
\end{eqnarray}
which allows to express a particle number conserving bilinear product of auxiliary
Fermi fields studied by Sch\"onhammer and Meden \cite{SchM96}

\begin{eqnarray}\label{1.29}
 \hat{\psi} _a (v) \equiv \sum^\infty _{l=-\infty} e^{ilv} \hat{c}_l
= \hat{\psi}_a (v + 2 \pi),
\end{eqnarray}
as

\begin{eqnarray}\label{1.30}
 \hat{\psi}^\dagger_a (u) \hat{\psi}_a (v) = G_N ( u - v )\,
\exp \left\{ - i \left[ \hat{\phi}^\dagger(u) - \hat{\phi}^\dagger(v) \right] \right\}
\exp \left\{ - i \left[ \hat{\phi} (u) -\hat{\phi} (v) \right] \right\}.
\end{eqnarray}
The quantity $G_N(u)$ in Eq. (\ref{1.30}) is a distribution valued Fermi sum defined by

\begin{eqnarray}\label{1.31}
\quad G_N (u) = \sum^{N-1} _{l = - \infty} e^{-il(u+i \epsilon)}.
\end{eqnarray}

In order to evaluate expectation values of exponentials containing
$\hat{\phi}$-operators, one must

\begin{itemize}
\item[a)] express the $\hat{d}$-operators in terms of the free $\hat{f}$ and
$\hat{f}^\dagger$-operators,
\item[b)] apply the bosonic Wick theorem.
\end{itemize}

The Wick theorem refers to homogeneous linear
combinations of $\hat{f}$ and $\hat{f}^\dagger$-operators. Due to backscattering,
the operator $\hat{\phi}$ contains a c-number part $\phi _c$, which must be
treated separately:

\begin{equation}\label{1.32}
\phi _c (u) \equiv - i C(-u),
\end{equation}
with

\begin{eqnarray}\label{1.34}
 C(u) \equiv \sum _{m=1}^\infty \xi _{2m}\,\exp(-2imu).
\end{eqnarray}
The quantities $\xi _{2m}$ are given by

\begin{eqnarray}\label{1.35}
\xi _{2m} \equiv \eta _{2m} \,\sqrt{\frac{K_{2m}}{2m}} =
 - V_c (m)\,\frac{K_{2m}}{4 m \, \epsilon _{2m} }.
\end{eqnarray}

Following the steps in \cite{WW01}, the zero temperature expectation value of
Eq. (\ref{1.30}) becomes

\begin{eqnarray}\label{1.36}
\langle \hat{\psi} _a ^\dagger (u) \hat{\psi} _a (v) \rangle = G_N (u-v) \,\exp[-W_1 (u,v)
- W_2 (u,v) ],
\end{eqnarray}
with $W_1$ given by

 \begin{eqnarray}\label{1.37}
 W_1 (u,v) =  \sum _{m>0} \frac{2}{m} [\gamma _m - \alpha _m \cos m (u+v) ]
 \left \{ 1 - \cos m (u-v) \right \}.
 \end{eqnarray}
The interaction parameters are

 \begin{equation}\label{1.38}
\alpha _m = \frac{1}{2} \sinh 2 \zeta _m, \quad \gamma _m = \sinh ^2 \zeta _m.
 \end{equation}
The contribution from the one-particle operator is:

\begin{eqnarray}\label{1.39}
W_2 (u,v) = C(-u)-C(u) +C(v)-C(-v)= 4i\sum _{m=1}^\infty \,\xi _{2m} \,\sin m(u-v)\,
\cos m(u+v).
\end{eqnarray}

In order to obtain the matrix elements, a number of calculational steps are done, starting
with the coordinate transformation $(u+v)/2 = t$,\, $u-v=s$. Using
$W_i(u,v)=\tilde{W}_i(s,t)$ and $ \tilde{W}_i (s \pm 2 \pi,t) = \tilde{W}_i (s,t \pm \pi)
= \tilde{W}_i(s,t)$, ($i=1,2$) and further symmetries, one obtains

 \begin{eqnarray}\label{1.40}
\langle \hat{c}^\dagger _{n-p} \hat{c} _{n+p} \rangle = \int^\pi _{-\pi} \frac{dt}{2 \pi} \cos pt\,
 \int^\pi _{-\pi}\frac{ds}{2\pi}\,
\exp [ - \tilde{W}_1(s,t/2)- \tilde{W}_2(s,t/2)]
\left\{\frac{\exp[i s(n-N+1)] }{1-\exp(is-\epsilon)}\right\}.
\end{eqnarray}
The distribution in curly brackets can be written as

\begin{eqnarray}\label{1.41}
\left\{\frac{\exp[is(n-N+1)]}{1-\exp(is -\epsilon)}\right\} = -
\left \{ \frac{\sin (n-N+ \frac{1}{2})s}{2 \sin(s/2)} \right \} +
i \left [ \frac{\cos (n-N + \frac{1}{2})s}{2 \sin(s/2)} \right ] +
\pi\, \delta _{2 \pi} (s).
\end{eqnarray}
By defining

\begin{equation}\label{1.42}
\tilde{W}_2(s,t/2) \equiv i f(s,t),
\end{equation}
the final result becomes

\begin{eqnarray}\label{1.43}
\langle \hat{c}^\dagger _{n-p}\hat{c} _{n+p} \rangle=\frac{1}{2}\,\delta _{p,0}
&-&\int _{-\pi}^\pi\,\frac{dt}{2\pi}\,\cos(pt)\,
\\[4mm]\nonumber
&&\times \int _{-\pi}^\pi
\,\frac{ds}{2\pi}\,\left\{\frac{\sin[(n-N+1/2)s-f(s,t)]}{2\sin(s/2)}\right\}
\exp[-\tilde{W_1}(s,t/2)].
\end{eqnarray}

Besides fixing the transformation parameters $\zeta _m$ in Eq.
(\ref{1.22}), the effect of backscattering  appears in the
argument of the sine via the function $f(s,t)$, which is
explicitly given by

\begin{equation}\label{1.44}
f(s,t) = 4 \sum _{m=1}\, \xi _{2m}\, \sin(m s)\, \cos(m t).
\end{equation}
Backscattering thus destroys the specific form of particle-hole symmetry found in \cite{WW01}
for forward scattering.

\section{Validity of the theory}

We can state that our solution Eq. (\ref{1.43}) is the exact
result for N one-dimensional fermions confined to a harmonic trap,
interacting via Eq. (\ref{1.9}), and immersed in an anomalous
vacuum of fermions filling all negative energy states. Thus we
have to assess the role of the anomalous vacuum for the present
finite size system. Intuitively, it is clear that its role
decreases with increasing Fermi energy $\epsilon _F \propto N$,
the energy region where interactions are most relevant.

We must, however, also consider the strength of the interaction.
The dominant dimensionless coupling constant in Eq. (\ref{1.27})
is $K_1 \equiv K$, since $K_m$ decreases with increasing $m$. It
is then seen that $K$ varies from zero to infinity when the
physical coupling coefficient $V_c(1)$ varies from $-\hbar \omega
_\ell$ to $\hbar \omega _\ell$. Values of $V_c(1)$ outside this
range are physically inaccessible, as are coupling constants
$g=g_2=g_4 < -\pi \hbar v_F $ in the corresponding Luttinger model
\cite{ML65,E79,V95,Sch95}. Note that at the extreme values
$V_c(1)= \pm \hbar \omega _\ell$ the renormalized excitation
energy $\epsilon _1$ according Eq. (\ref{1.25}) vanishes.

In the Luttinger model, $K \rightarrow \infty$ or $g \rightarrow
-\pi \hbar v_F$ corresponds to the strongest physically allowed
attraction. The compressibility vanishes and a phase separation
occurs (cf. the discussion in \cite{V95}).

Our numerical results for the particle density in the present model show an increasing
extension beyond the classical
turning points, i.e., the density progressively leaks out of the trap for increasing
$K \gg 1$. The interpretation of this effect is difficult due to the presence of the
anomalous vacuum and the finiteness of $N$: It is conceivable that an increasing
interaction pulls more and more fermions out of the anomalous vacuum as
indicated by studies of the fermion sum rule below.

Nevertheless, we can make the following statement about the
validity of our bosonization scheme: For any fixed $|V_c|/(\hbar
\omega _\ell) < 1$, i.e., $0<K<\infty$, the error due to the
anomalous vacuum can be made as small as wanted by increasing the
physical particle number $N$. We call this asymptotically exact.

Though we do not have an analytic expression for the error at present, we know
from our study of the fermion sum rule and the results presented below that the error
decays fast with increasing $N$, probably in an exponential way.

The precise nature of the singular points $K=0$ and $K= \infty$ in
the present finite system requires, however, further
investigation. In assessing its relevance for interacting confined
fermions, one must also consider the dependence of $V_c$ on the
particle number. This in turn depends on the specific form of the
interaction potential (cf. \cite{GW02}). It thus seems that there
is no simple and general limit $N \rightarrow \infty $ accompanied
by a proper scaling of $\omega_l$ and the interaction parameters
("thermodynamic limit") for the interacting system.

We expect, however, that the region near the center of the trap
($|z| \ll L_F \propto \sqrt{N}$) acquires properties of a
homogeneous Luttinger liquid. This is demonstrated for the central
pair correlation function studied in the following Section and for
a reasonably large particle number $N=10^3$.

We also note that numerical investigation of the interaction coefficients in \cite{GW02}
show that increasing $N$ does not alter the dominance of $V_c$ over the forward scattering
coefficients in a gas of identical fermions.

\subsection{Numerical Method}

For the purpose of the present investigation, it is sufficient to
use a simplified model ("toy" model IM1 in \cite{WW01}) by
retaining only the terms with $v= \pm 1$ in Eq. (\ref{1.9}). The
relevant parameter then are:

\begin{eqnarray}\label{3.1}
\alpha _{1} =  \frac{1-K^2}{4 K},\quad \gamma _1 = \frac{(1-K)^2}{4 K}.
\end{eqnarray}

The function $\tilde{W_1}(s,t/2)$ in Eq. (\ref{1.43}) becomes $2[
\gamma _1-\alpha _1 \cos(t)][1-\cos(s)]$ and the backscattering
expression (\ref{1.44}) simplifies to $f(s,t) = -
V_c(1)\,\sin(s)\,\cos(t)/\hbar \omega _\ell$. The one-particle
matrix elements at zero temperature are then given by

\begin{eqnarray}\label{3.2}
\langle \hat{c}^ \dagger _{M-p}\hat{c} _{M+p} \rangle
=\frac{1}{2}\,\delta _{p,0} -\int _\pi ^\pi \frac{dt}{2
\pi}\,\cos(pt)&&\int _{-\pi}^\pi \frac{ds}{2 \pi}
\left\{\frac{\sin[(M+1/2-N)s - f(s,t)]}{2 \sin(s/2)}\right\}
\\[4mm]\nonumber
&&\times \exp\left\{-2[ \gamma _1-\alpha _1
\cos(t)][1-\cos(s)]\right\}.
\end{eqnarray}

This is the prediction of the bosonization method for identical fermions in the one-dimensional
harmonic trap with dominant backscattering.

In \cite{WW01}, it was already pointed out that first order
perturbation theory in fermionic Hilbert space reproduces the
results using the equation corresponding to Eq. (\ref{3.2}) for
forward scattering ($f$=0) when it is expanded to first order in
$\alpha _1$ and $\beta _1$, and the same applies to the present
case: Both approaches lead to the weak coupling result

\begin{eqnarray}\label{3.2a}
\langle \hat{c}^ \dagger _{M-p}\hat{c} _{M+p} \rangle =\delta
_{p,0}\,\Theta(N-M-1/2) -\left(\frac{V_c(1)}{2 \hbar \omega
_\ell}\right)\,\delta _{M,N-1}\,(\delta _{p,1} +\delta _{p,-1}) +
O \left(\frac{V_c(1)}{\hbar \omega _\ell }\right)^2.
\end{eqnarray}

Finally, the predictions of Eq. (\ref{3.2}) will be checked
against numerical results obtained by direct diagonalization in
the fermionic $N$-particle Hilbert space and for strong coupling.

The fermionic $N$-particle Hilbert space is spanned by the unperturbed $N$-particle states

\begin{equation}\label{3.3}
|\{m\}\rangle^{(0)}=\hat{c}_{m _{1}}^  {\dagger} \hat{c}_{m _{2}}^
{\dagger} \cdots  \hat{c}_{m _{N}}^  {\dagger}
 |vac \rangle.
\end{equation}
Here, $\{m\}$ denotes a sequence of occupation numbers $m_n=0,1$
for the single particle states $\psi _n(z)$. For example, the
three-particle state with energy $(11/2)\,\hbar \omega _\ell$ and
excitation energy $\hbar \omega _\ell$ is

\begin{equation}\label{3.4}
|1,1,0,1,0,0,\ldots \rangle^{(0)} = \hat{c}_{0}^  {\dagger}
\hat{c}_{1}^  {\dagger} \hat{c}_{3}^  {\dagger} |vac \rangle.
\end{equation}

In order to simplify the notation for the unperturbed $N$-particle states, we classify
them according to their excitation energies $\Delta E(n)= n \hbar \omega _\ell$. Degeneracy
is taken care of by ordering the states
according to the lowest unoccupied single particle state occurring in them. They are then
numbered consecutively $|m \rangle^{(0)}, m=0,1,2,\ldots$. The unperturbed ground state is thus

\begin{equation}\label{3.5}
|0 \rangle^{(0)}  = |1,1,\ldots,1,1,0,0,\ldots\rangle,
\end{equation}
and subsequent excited states are

\begin{eqnarray}\label{3.5a}
|1 \rangle^{(0)} &=& |1,1,\ldots,1,0,1,0,0,\ldots\rangle,
\\[4mm]\nonumber
|2 \rangle^{(0)} &=& |1,1,\ldots,0,1,1,0,0,\ldots\rangle,
\\[4mm]\nonumber
|3 \rangle^{(0)} &=& |1,1,\ldots,1,0,0,1,0,0,\ldots\rangle,\quad \mbox{etc.}.
\end{eqnarray}

The perturbed ground state to first order used to obtain Eq.
(\ref{3.2a}) is

\begin{equation}\label{3.5b}
|0 \rangle^{(1)}  = |0 \rangle^{(0)} +\left(\frac{V_c(1)}{2 \hbar \omega _\ell}\right)
\,|2 \rangle^{(0)}.
\end{equation}
This result was also used to check the numerical procedure.

The actual eigenstates of the interacting Hamiltonian are denoted
by $|s \rangle$. They are expanded according to

\begin{equation}\label{3.6}
|s \rangle = \sum _m \,c_{sm}\, |m \rangle^{(0)}.
\end{equation}
These expansion coefficients are the central quantities in the
numerical procedure: In terms of the expansion coefficients, the
occupation probabilities for oscillator states $\psi _M$ with
respect to the ground state $|0 \rangle$ are

\begin{equation}\label{3.7}
P(M)=\sum _{mn} c_{0m} c_{0n}^{*}\, ^{(0)}\langle n | \hat{c}_{M}^
{\dagger}\hat{c}_{M} |m \rangle ^{(0)}.
\end{equation}

Similarly, the particle density, which is the expectation value of
the density operator $\hat{\psi}^ \dagger(z)\hat{\psi}(z)$ with
$\hat{\psi}(z)\equiv \sum _n \psi _n(z) \,\hat{c}_n$, becomes

\begin{eqnarray}\label{3.8}
n(z) = \langle 0|\hat{\psi}^ \dagger(z)\hat{\psi}(z)|0 \rangle =
\sum _{p=0,q=0}^\infty \psi _{p}(z)\psi _{q}(z)\, \sum _{mn}\,
c_{0m}\, c_{0n}^{*} \,^{(0)}\langle n|\hat{c}_{p}^
{\dagger}\hat{c}_{q} |m \rangle^{(0)}.
\end{eqnarray}

The first computational step is the evaluation of the interaction matrix elements

\begin{equation}\label{3.9}
 V_{ij}=\, ^{(0)}\!\langle i|\hat{V}_c|j \rangle^{(0)} .
\end{equation}
In detail, the free $N$-particle eigenstates Eq. (\ref{3.3}) and
the expectation values in Eqs.(\ref{3.7}), (\ref{3.8}), and
(\ref{3.9}) with respect to these states are calculated via an
algorithm which implements the fermion-algebra. The number of
eigenstates which are taken into account grows strongly with the
maximal excitation energy $\Delta E(n_{max})$ due to increasing
degeneracy.

In subsequent steps, the diagonalization of the matrix

\begin{equation}\label{3.10}
H_{ij}={H_{0}}_{ij} + V_{ij},
\end{equation}
and the computation of the expansion coefficients $c_{sm}$ follows
\cite{FL00,LG93}. Eigenvalues and eigenvectors of the matrix
(\ref{3.10}) are computed using a QL algorithm. The normalized
eigenvector of the ground state determines the required expansion
coefficients $c_{0 m}$ according to $^{(0)}\langle m|0 \rangle$.

\subsection{Occupation Probabilities}

The occupation probabilities of oscillator states $\psi _M$ in the interacting ground
state are

\begin{eqnarray}\label{3.11}
0 \le P(M)=\langle 0|\hat{c}^ \dagger _M \hat{c} _M |0 \rangle \le
1.
\end{eqnarray}
Due to the backward scattering, they do not show the symmetry
found in \cite{WW01,GaW02} for forward scattering.

We will also discuss the sum rule

\begin{eqnarray}\label{3.12}
S(N,\alpha _1) \equiv\sum _{m=0}^\infty \,\langle \hat{c}^ \dagger
_m
 \hat{c} _m  \rangle\stackrel{?}{=} N
\end{eqnarray}
for the fermion number.
This sum rule is only asymptotically fulfilled in the bosonization method:
The particles in the anomalous vacuum
couple to the physical particles, giving an effective particle number $S(N,\alpha _1)
>N$ \cite{WWW}. The effect is most pronounced when the particle number $N$ is smaller than
the coupling strength $|\alpha _1|$. The coupling to the anomalous vacuum also leads to
the surprising feature that interaction effects occur for just one physical particle ($N=1$).

\begin{center}
\includegraphics[width=9cm,angle=270]{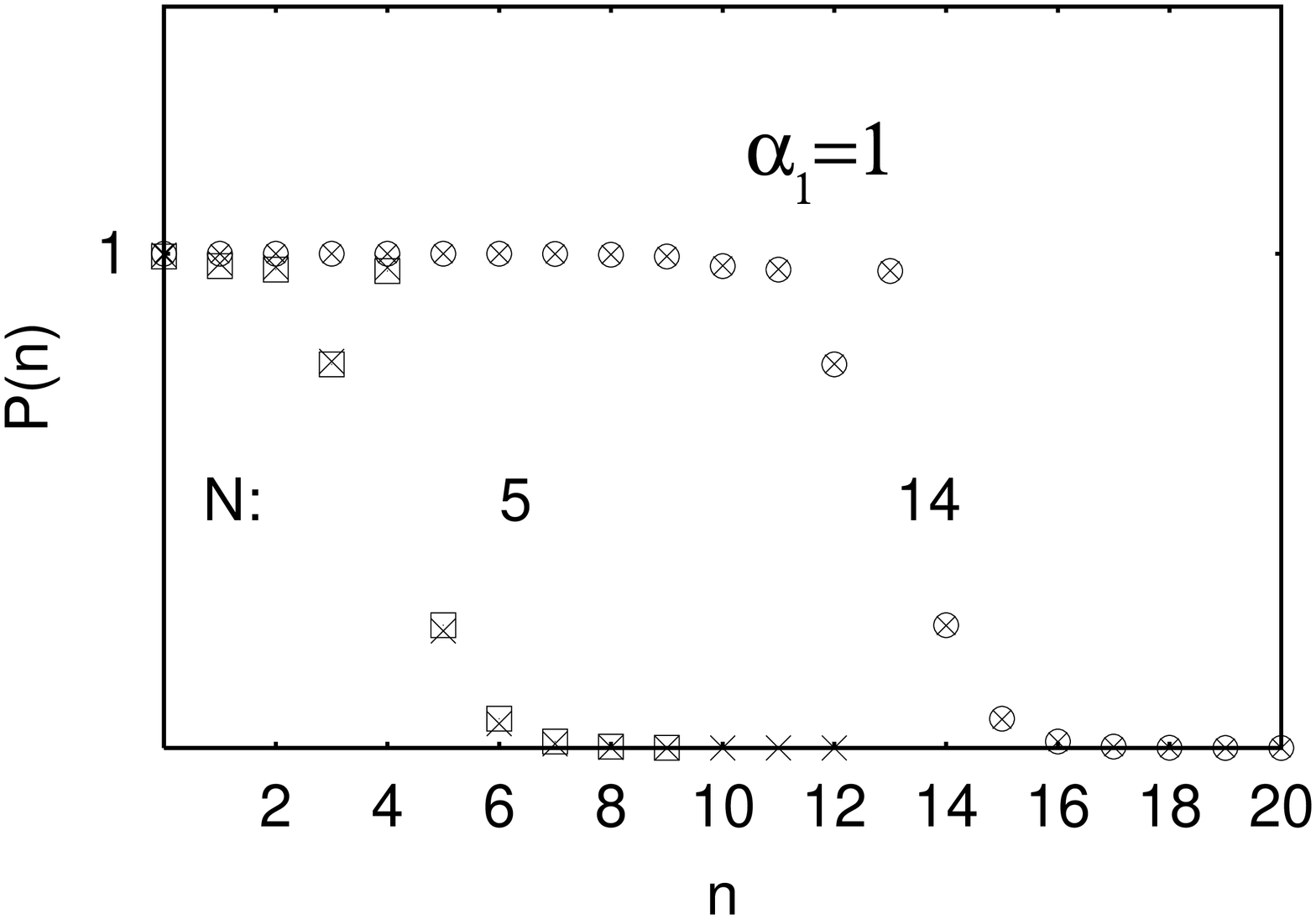}
\end{center}
\begin{description}
\item[Fig. 1:] Occupation probabilities $P(n)$ of oscillator states $\psi _n$ in the
interacting system for repulsive backscattering strength $\alpha
_1=1$ at zero temperature and particle numbers $N=5$ and $N=14$.
Crosses are results of numerical diagonalization, predictions of
the bosonization method are given as squares ($N=5$) and circles
($N=14$), respectively. No significant deviations between the two
approaches are seen for particle number $N=14$. For $N=5$, the
presence of the anomalous vacuum in the bosonization method causes
visible deviations.
\end{description}

Figure 1 shows a comparison of occupation probabilities calculated
using the bosonization method and numerical diagonalization. For a
coupling constant $\alpha _1=1$, which corresponds to $V_c(1)=
-0.894 \,\hbar \omega _\ell$,  no significant deviations are found
for particle numbers $N=14$ and larger. However, for $N=5$, the
bosonization method is less accurate because the anomalous vacuum
is present. Though the excess $\Delta N(5,1) \equiv S(5,1)-N$
amounts to only $5.4 \cdot 10^{-4}$, individual occupation
probabilities are less accurate.

\subsection{Particle density}

The density of the trapped particles can be written as

\begin{eqnarray}\label{3.13}
n (z) = \sum^\infty _{M = 0} \psi _{M}(z)^2\,\langle \hat{c}^
\dagger _M \hat{c} _M \rangle +2 \sum^\infty _{M = 1} \sum
_{p=1}^{M}\psi _{M-p}(z)\,\psi _{M+p}(z)\, \langle \hat{c}^
\dagger _{M-p}\hat{c} _{M+p} \rangle.
\end{eqnarray}

The particle density shows Friedel oscillations \cite{GWSZ00,F58}.
The amplitudes of the Friedel oscillations are modified by the interaction
as described in \cite{WW01,GaW02,GSchW01}.

Figure 2 displays particle densities for the coupling value
$\alpha _1=10$, corresponding to $V_c(1)=-0.998 \,\hbar \omega
_\ell$, which is a rather large repulsive interaction. The results
of numerical diagonalization and of the bosonization method are
compared for various particle numbers $N$. The deviations found
are due to the presence of the anomalous vacuum in the
bosonization method which interacts with the physical particles.
The differences are more pronounced for larger coupling strength
or smaller particle number $N$: Larger particle numbers lead to
larger Fermi energies $\epsilon _F $. This decouples the
physically relevant energy region $\epsilon \approx \epsilon _F$
from the anomalous vacuum. Larger coupling strengths counteract
the decoupling.

\begin{center}
\includegraphics[width=10cm,angle=270]{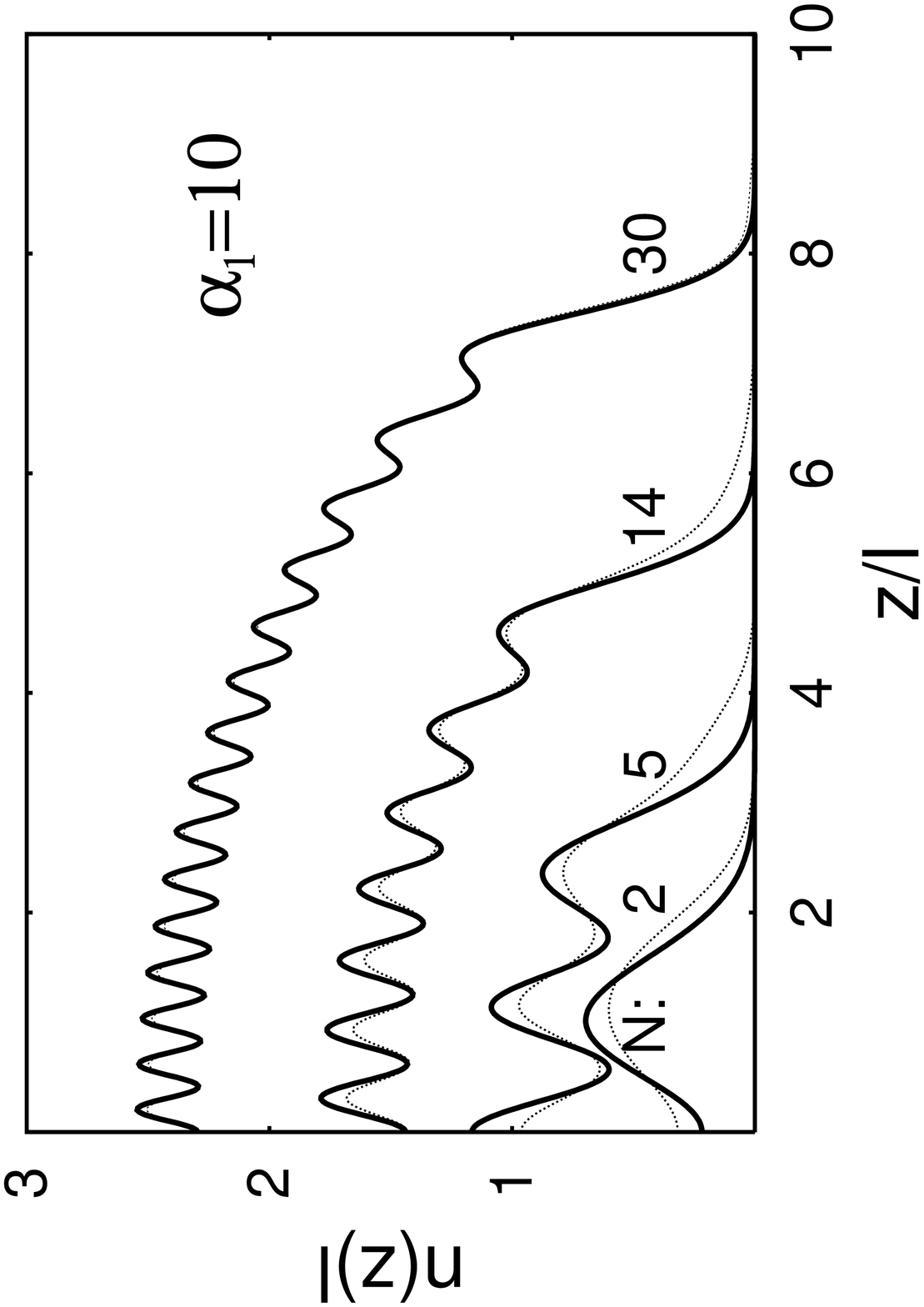}
\end{center}
\begin{description}
\item[Fig. 2:] Dimensionless particle density $n(z) l$ in units of the inverse of the
 oscillator length $l$ versus dimensionless distance $z/l$ from the center of the
 one-dimensional harmonic
 trap for several particle numbers $N$ of a Fermi gas with dominant backward scattering
 at zero temperature.
 Thick curves are from numerical diagonalization and
 dotted curves show the Friedel oscillations calculated by the bosonization
 method. Interaction strength is $\alpha _1=10$. Deviations for particle numbers
 less than $N=30$ are due to the anomalous  vacuum, which couples strongly to the
 real particles.
\end{description}

We can summarize: The results of the bosonization method for the
one-particle matrix elements of a harmonically trapped
one-dimensional Fermi gas with backscattering when compared with
the exact numerical diagonalization show noticeable deviations
from the exact results when the physical number $N$ of fermions is
small or the interaction strength is large. This effect is not
restricted to backscattering. Forward scattering would produce
similar results. For sufficiently large $N$, the bosonization
method produces, however, fully acceptable results. It is then far
more effective than any numerical approach.

\section{Results for large N}

The functions $\tilde{W}_1(s,t)$ and  $f(s,t)$ in the exact expression (\ref{1.43}) for
the matrix
elements contain a bewildering number of interaction coefficients, which all depend
on the index $m$. In order to proceed with the evaluation, we describe in the
Appendix a model for the interaction coefficients, which allows to do all
summations and express them by just two parameters: The main coupling constant $K$
and a small number $r \ll 1$ specifying the exponential decay of all interaction
coefficients. Following \cite{GaW02}, we adopt $r=1/\sqrt{N}$ so that
only the coupling constant $K>0$ remains. $K>1$ corresponds to attractive interactions and
$0<K<1$ to repulsion between the fermions. An analogous procedure is used in the theory
of the Luttinger model \cite{E79,Haldane,V95,Sch95}.

The result for the one-particle matrix elements becomes

\begin{eqnarray}\label{2.1}
\langle \hat{c}^\dagger _{n-p}\hat{c} _{n+p} \rangle= \frac{\delta _{p,0}}{2}
&-&\int^{\pi}_{-\pi}\frac{d t}{2 \pi}\, \frac{\cos(p\,t)}{[1+Z_\alpha-\cos(t)]^{\alpha _0}}
\int^{\pi}_{-\pi} \frac{ds}{2 \pi}\,\left\{\frac{\sin[(n-N+1/2)s-f(s,t)]}{2 \sin(s/2)}\right\}
\\[4mm]\nonumber
&&\times\left\{\frac{Z_\gamma}{[1+Z_\gamma-\cos(s)]}\right\}^{\gamma _0}
\left\{[1+Z_\alpha-\cos(t-s)][1+Z_\alpha-\cos(t+s)]\right\}^{\alpha _0/2},
\end{eqnarray}
with constants $Z_\gamma$ and $Z_\alpha$ according to

\begin{eqnarray}\label{2.2}
Z_\gamma=\cosh(r_\gamma)-1 \rightarrow r^2/2,\quad Z_\alpha=\cosh(r_\alpha/2)-1
\rightarrow r^2/8.
\end{eqnarray}
The coupling dependent exponents in Eq. (\ref{2.1}) are

\begin{eqnarray}\label{2.3}
\alpha _0=\frac{1}{4K}\,(1-K^2),\quad \gamma _0 = \frac{1}{4K}\,(1-K)^2 \ge 0.
\end{eqnarray}
The sign of $\alpha _0$ is negative when the backscattering is attractive.

In addition, we take advantage of the fact that backscattering is
dominant in a harmonically confined gas of identical fermions: The
function $f$ is then determined by the backscattering contribution
$C(u)$ in Eq. (\ref{4.14}):

\begin{eqnarray}\label{2.4}
f(s,t)= 2 \Im[C(t/2-s/2)-C(t/2+s/2)]
= \frac{1}{4}\,(1-K^2)\, \arctan \left [ \frac{2\,q\,
(\cos t - q\, \cos s)\,\sin s}{1 -q^2 - 2 \,q\,(\cos t -q\,\cos s)\,\cos s}
\right ],
\end{eqnarray}
with

\begin{eqnarray}\label{2.4a}
q=\exp(-r/2)\rightarrow 1-r/2+r^2/8.
\end{eqnarray}

\subsection{Particle and momentum densities}

The particle density $n(z)$ is evaluated using Eq. (\ref{3.13})
and the matrix elements Eq. (\ref{2.1}). For large $N$ and
moderate coupling ($K$ is of order unity), the particle density
well inside the trap is dominated by the large number of nearly
filled states inside the Fermi sea. However, the states above the
Fermi level $N_F=N-1$ make significant contributions to $n(z)$ by
modifying the Friedel oscillations \cite{F58}: This is
demonstrated in Fig. 3 using the full expression (\ref{2.1}).

 \begin{center}
 \epsfig{figure=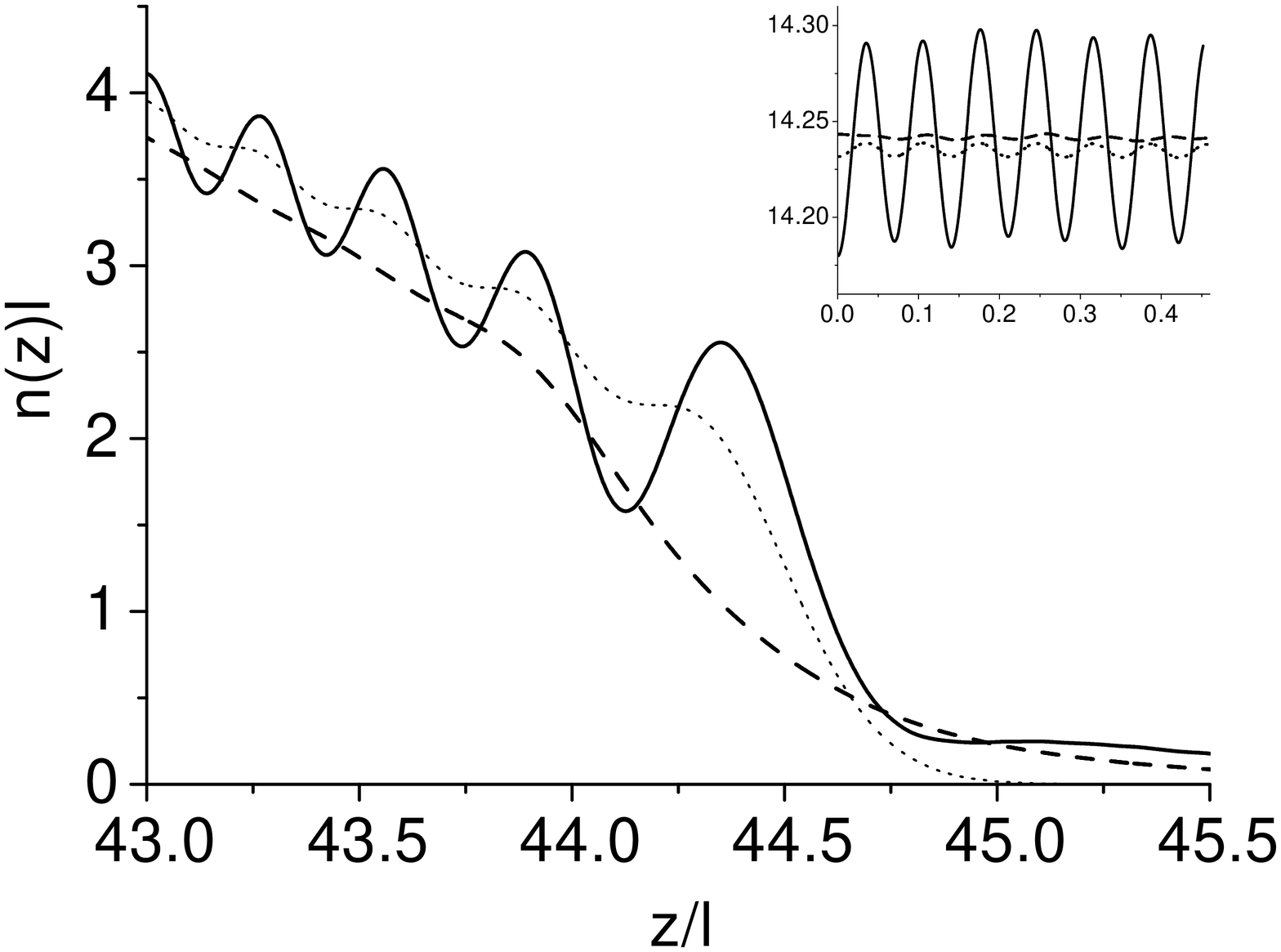,width=14cm}
 \end{center}
 \begin{description}
 \item[Fig. 3:] Dimensionless particle density $n(z) l$ in units of the inverse of
 the oscillator length $l$
 versus dimensionless distance $z/l$ near the classical boundary at $L_F=l \sqrt{2N-1}$ of
 the one-dimensional harmonic trap for $N=1000$ interacting spinless fermions at zero
 temperature. Dotted curve shows unperturbed Friedel oscillations. Strongly oscillating
 curve refers to a repulsive interaction with $K=1/3$ while the dashed curve
 is for an attractive interaction with $K=3$. The inset displays the density near the center
 of the trap, where Friedel oscillation have the usual period $\pi/k_F
 =\pi l/\sqrt{2N-1}$.
\end{description}

It is seen that Friedel oscillations are enhanced in the repulsive
case while an attractive interaction suppresses them. Near the
classical boundary $|z| \approx L_F = l\,\sqrt{2N-1}$, the
oscillation period is larger than the standard value $\pi/k_F$
well inside the trap and obeys the relations found in
\cite{GWSZ00} for the non-interacting case. Furthermore, it is
seen that the perturbed density extends considerably into the
classically forbidden region. The effect becomes stronger with
increasing coupling strength ($ K \gg 1$ or $\ll 1$).

The momentum density distribution is given by

\begin{eqnarray}\label{2.14}
p(k)=l^2\,\sum _{n = 0}^\infty\sum _{p=-n}^n (-1)^p \,\psi _{n-p}
(k \,l^2)\,\psi _{n+p} (k\,l^2)\,\langle \hat{c}^\dagger_{n-p}\hat{c}_{n+p}\rangle.
\end{eqnarray}

It is seen that $p(k)$ is identical in shape to the density distribution $n(z)$
provided the diagonal approximation for the one-particle matrix elements is
reasonable. This is the case for moderate attractive coupling.

We know \cite{WW01} that the Friedel
oscillations in $n(z)$ and $p(k)$ behave oppositely: Attractive interactions
increase them in the momentum density and decrease them in the particle density and
vice versa for repulsion. A corresponding effect is seen in the Wigner function
discussed below.

\subsection{Wigner function}

In terms of the local creation and annihilation operators $\hat{\psi} ^\dagger
(z)$ and $\hat{\psi}(z)$, the static Wigner function
of the many-fermion system is given (cf. \cite{HOSW84}) by

\begin{eqnarray}\label{2.15}
W(z,k)  =  \int ^\infty _{-\infty} \, d \zeta\, e^{-i k \zeta} \langle \hat{\psi} ^\dagger
(z-\zeta/2)\, \hat{\psi}(z+ \zeta/2)\rangle.
\end{eqnarray}
The Wigner function of the one-dimensional Fermi gas with two components and
forward scattering between the two components was studied in \cite{GSchW01}.

Transforming to the oscillator representation gives

\begin{eqnarray}\label{2.16}
W(z,k)=\sum^\infty _{m,n=0}  \langle \hat{c} ^\dagger _{m} \hat{c}_{n} \rangle
\,f_{mn} (z,k),
\end{eqnarray}
with expansion coefficients $f_{mn}$ according to

\begin{eqnarray}\label{2.17}
 f_{mn} (z,k)  =   \int^\infty _{-\infty} d \zeta \,e^{-i k\zeta} \,\psi _m (z- \zeta/2)
\,\psi _n (z + \zeta/2)=f_{nm} (z,-k).
\end{eqnarray}
These are explicitly given in terms of generalized Laguerre
polynomials ($n \geq m$, cf. e.g., \cite{Sch01}) by

\begin{eqnarray}\label{2.18}
f_{mn} (z,k)  = 2  (-1) ^m \,(2^{n-m} m!/ n!)^{1/2}\, (z/l -
ikl)^{n-m}\, \exp \left(- z^2/l^2 - k^2 l^2 \right) {\rm
L}^{(n-m)}_m \left(2 z^2/l^2 +2 k^2 l^2 \right).
\end{eqnarray}
It is noted that the one-particle matrix elements can be completely reconstructed from the
Wigner function:

\begin{eqnarray}\label{2.19}
\langle \hat{c} ^\dagger _{m} \hat{c}_{n} \rangle = \frac{1}{2 \pi} \int ^\infty _{-\infty}
\, d z\, d k\, \,f_{nm}(z,k)\,W(z,k).
\end{eqnarray}
The Wigner function is thus equivalent to the full set of one-particle matrix elements.

We show an example of the Wigner function in Figs. 4(a) and 4(b)
for a repulsive interaction.

\begin{center}
 \epsfig{figure=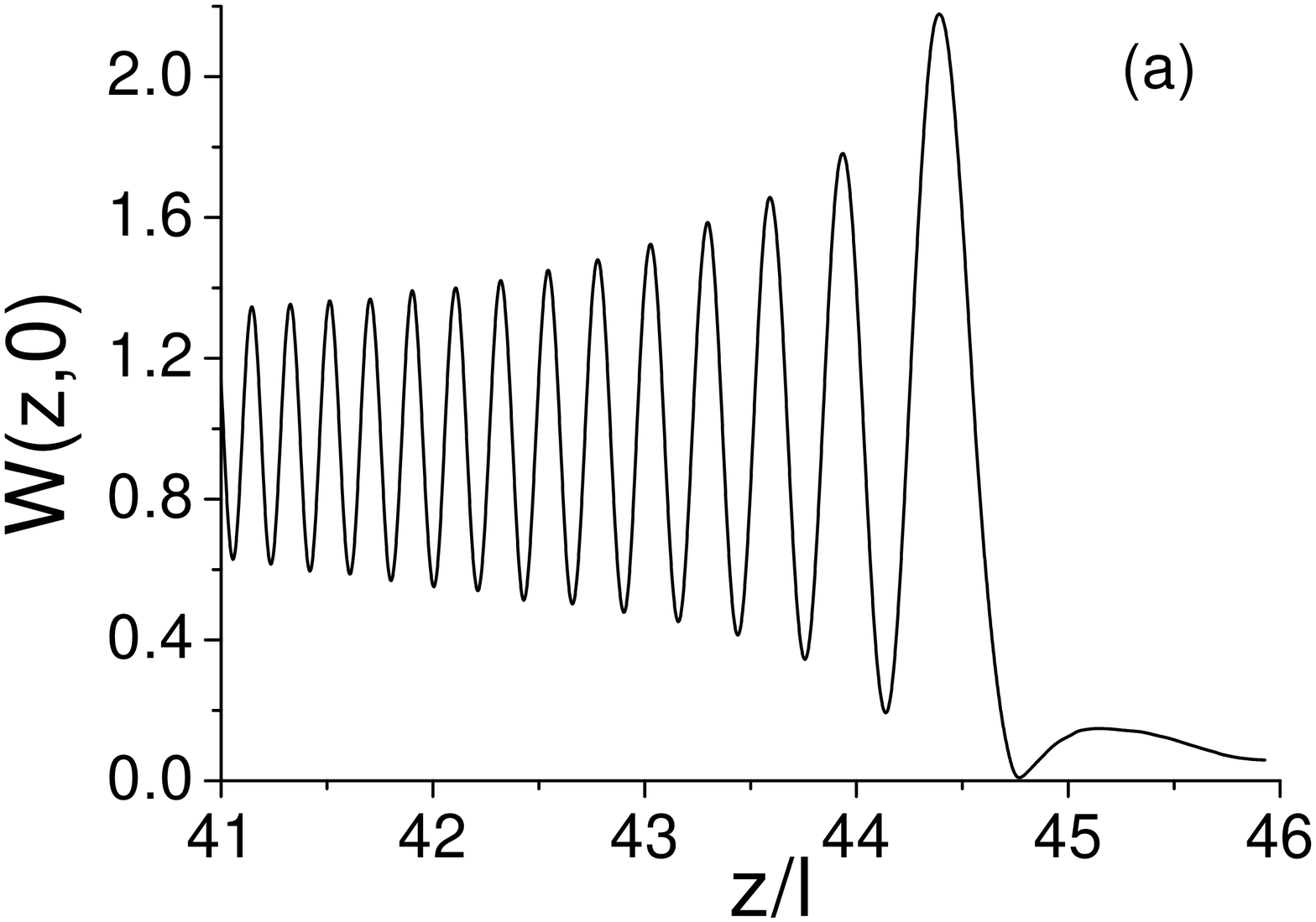,width=8cm}
  \epsfig{figure=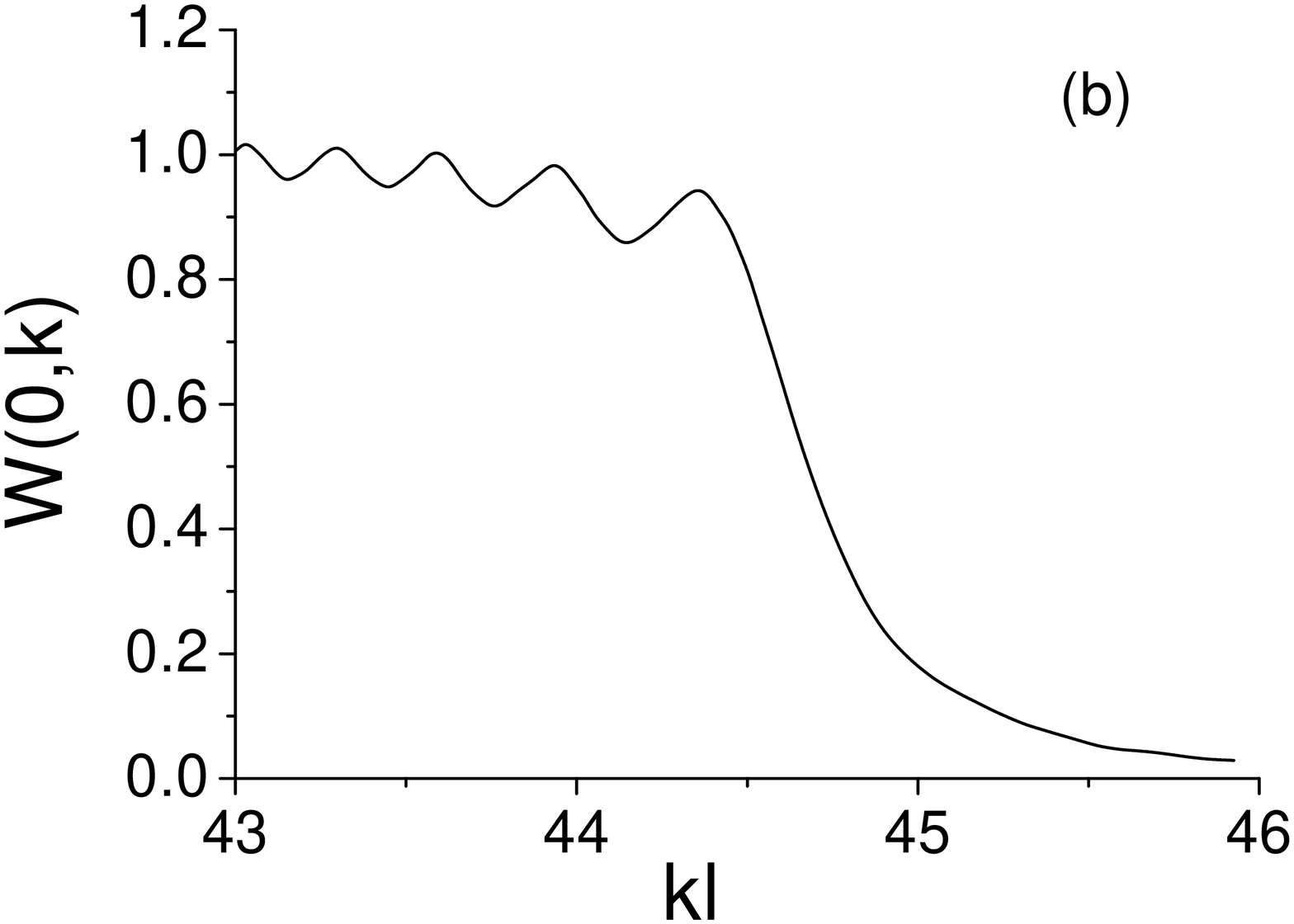,width=8cm}
 \end{center}
\begin{description}
\item[Fig. 4:] Friedel oscillations in phase space: Wigner function for $N=1000$ interacting
 spinless fermions at zero temperature in the one-dimensional harmonic trap. In (a), the
 cross section
 $W(z,k=0)$ is plotted versus dimensionless distance $z/l$ displaying the region
 around the classical turning point $L_F=l\sqrt{2N-1}$. l is the oscillator length.
 In (b), the corresponding function
 $W(z=0,k)$ is shown versus dimensionless momentum $kl$. Note the increase of the
 oscillation amplitude near the classical boundary. The plot refers to a repulsive
 interaction with $K=1/3$. Repulsive interactions enhance Friedel oscillations in the
 spatial direction and suppress them in the momentum direction.
 \end{description}

The amplitudes of the Friedel oscillations increase near the
classical turning point. This was discussed for the
non-interacting case \cite{GWSZ00}.

Neglecting non-diagonal matrix elements in Eq. (\ref{2.16}) would
give $W(z,k=0)=W(z=0,k \rightarrow z/l^2)$. The significant
differences between Fig. 4(a) and Fig. 4(b) thus stem from the
non-diagonal matrix elements, which are particular relevant for
repulsive interactions.

The static pair correlation function with respect to the center of the trap

\begin{eqnarray}\label{2.22}
C(z,z'=0) \equiv \langle \hat{\psi} ^\dagger(z)\, \hat{\psi}(z'=0)\rangle=
\frac{1}{2 \pi} \int ^\infty _{-\infty} \, d k\, e^{-i kz}\,W(z/2,k),
\end{eqnarray}
as well as any other one-particle property of the many-particle system are also contained
in the Wigner function.

The central pair correlation function (\ref{2.22}) is displayed in Fig. 5. It is noted that
the wave length of the intrinsic periodicity in the center of the trap is twice that of the
Friedel oscillations, i.e.,
$\lambda = 2 \pi/k_F$. In fact, the central pair correlation function for non-interacting
fermions is given by

\begin{eqnarray}\label{2.23}
C_0(z,z'=0) = \frac{1}{l\,\sqrt{\pi}}\,\,e^{- z^2/(2\,l^2)}\,\left[\sum _{n=0}^M (-1)^n\,
\frac{H_{2n}(z/l)}{4^n n!}\right]
\rightarrow \frac{\sin(k_F z)}{\pi z},
\end{eqnarray}
provided $N=2M$ is large and $z$ is restricted to the center of the trap ($|z| \ll L_F$).
We expect that interactions modify Eq. (\ref{2.23}) according to

\begin{eqnarray}\label{2.24}
C(l \ll z \ll L_F,z'=0)  \propto \frac{\sin(k_F z)}{l\,(z/l)^{\alpha _C}},
\end{eqnarray}
i.e., the anomalous dimension of the correlation function is $\alpha _C =1+2 \gamma _0
=(K+1/K)/2$,
as in a Luttinger liquid. This is confirmed numerically by displaying the corresponding
envelope in Fig. 5.

 \begin{center}
 \epsfig{figure=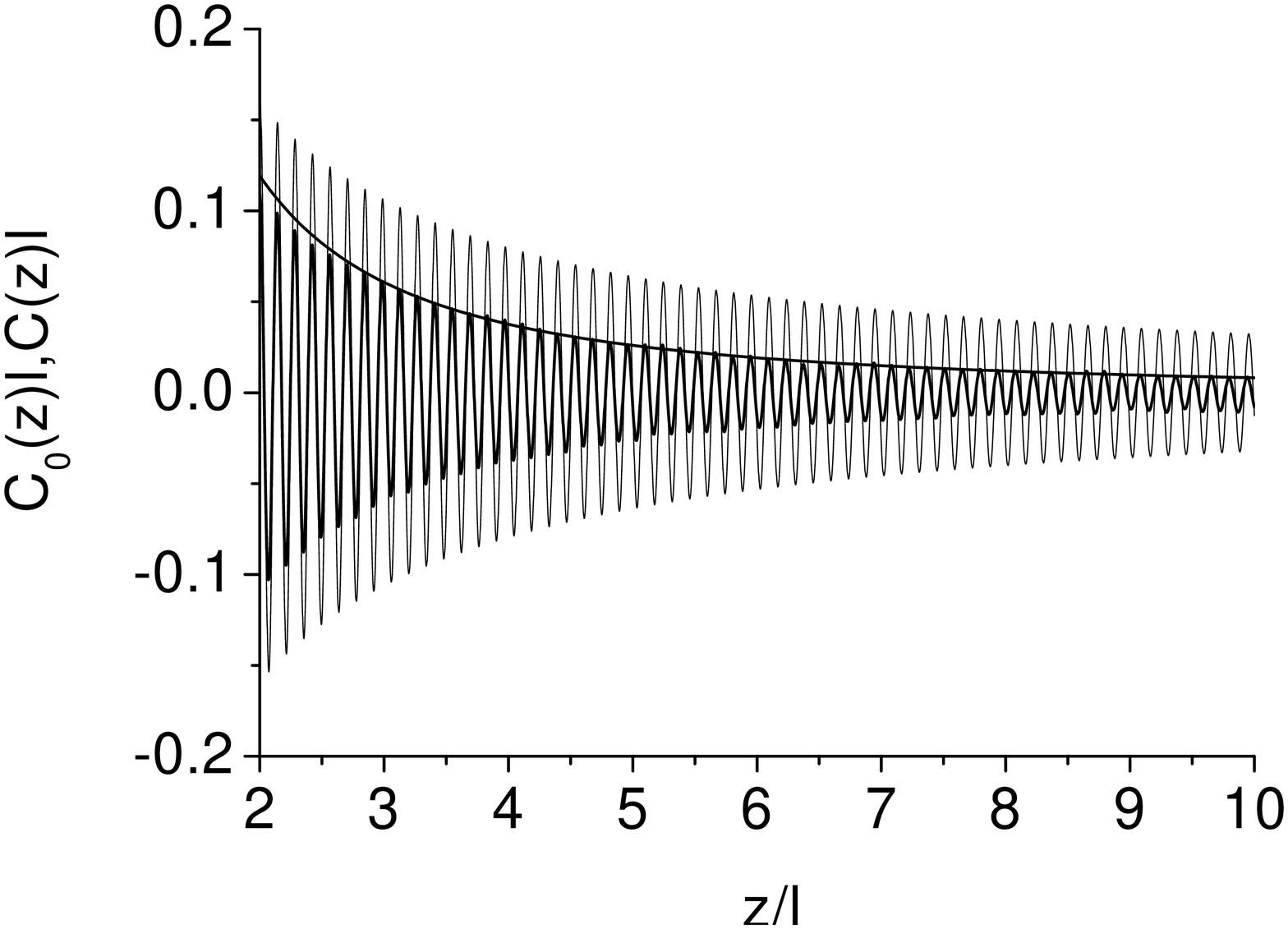,width=14cm}
 \end{center}
 \begin{description}
 \item[Fig. 5:] Dimensionless correlation functions $C_0(z,z'=0)l$ and $C(z,z'=0)l$ in units
 of the inverse of the oscillator length $l$
 versus dimensionless distance $z/l$ from the center of the trap for $N=1000$ spinless
 fermions at zero temperature. Thin curve shows non-interacting function $C_0$ while
 the thick curve gives $C$ for an attractive interaction with $K=3$. Oscillations have twice
 the period of Friedel oscillations. Envelope to the full curve displays power law decay
 with anomalous dimension $(K+1/K)/2$.
\end{description}

The correlation function $C$ of the interacting system thus decays
faster than $C_0$ and this effect is invariant under $K
\rightarrow 1/K$.

\section{Summary}

The fact that backscattering dominates the interaction between
identical one-dimensional fermions confined to a harmonic trap
makes it necessary to include backscattering into the bosonization
method. This can be done exactly by supplementing the squeezing
transformation (\ref{1.18}) with the displacement transformation
(\ref{1.20}). This changes (in fact complicates) the result
(\ref{1.43}) for the one-particle matrix elements by altering the
argument in the sine-function. This modification destroys the
symmetry

\begin{eqnarray}\label{2.25}
\langle \hat{c}^\dagger _{2N-1-n-p}\hat{c} _{2N-1-n+p} \rangle
= \delta _{p,0}-\langle \hat{c}^\dagger _{n-p}\hat{c} _{n+p} \rangle.
\end{eqnarray}

In order to evaluate the new result, we introduce in analogy to
the Luttinger model a simplified form for the interaction
coefficients which uses only one coupling constant $K$. $K=1$
corresponds to the non-interacting case studied in
\cite{GWSZ00,VMT00,MVT01,VMT01}, $K >1$ corresponds to attraction.
For values of $K$ near unity, the Fermi edge at zero temperature
is already significantly smoothed out by the interactions, an
effect which becomes more pronounced for stronger coupling.

We then study particle and momentum densities for various coupling strengths. Friedel
oscillations in phase space are characteristic for these quantities, which extend
progressively into the classically forbidden region when the coupling strength is
increased. The results confirm an earlier finding that attractive interactions
decrease the amplitude of the Friedel oscillations in real space while repulsion enhances
it. The Friedel oscillations in momentum space behave oppositely.

Finally, the central pair correlation function is calculated. The basic period of the pair
correlation function in the center of the trap is $2 \pi/k_F$ in contrast to $\pi/k_F$ of
the Friedel oscillations. Interactions do not affect these basic periodicities: Even the
increase of the Friedel period in the particle density near the classical boundaries is
correctly given by the non-interacting theory. Interactions, however, modify the power
law decay of the correlation function inside the trap. Our numerical results are consistent
with the value

\begin{eqnarray}\label{2.26}
\alpha _C = \frac{K+1/K}{2}
\end{eqnarray}
for the anomalous dimension $\alpha _C$ of the correlation well inside the trap, in accordance
with the predictions for a Luttinger liquid.

We also assessed the validity of the bosonization method, comparing its result with those of
direct numerical diagonalization in fermionic Hilbert space.

\begin{acknowledgments}
We gratefully acknowledge helpful discussions with G. Alber, T.
Pfau, and W. P. Schleich and financial help by Deutsche
Forschungsgemeinschaft.
\end{acknowledgments}

\section{Appendix}

\newcounter{affix}
\setcounter{equation}{0}
\setcounter{affix}{1}
\renewcommand{\theequation}{\Alph{affix}\arabic{equation}}

\section*{\bf A model for the interaction coefficients}

The result (\ref{1.43}) for the one-particle matrix elements
contains three sets of interactions coefficients: $\{\alpha _m
\}$, $\{\gamma _m\}$, and $\{V_c(m) \}$. The first two sets depend
on the basic interaction coefficients $V_c(m)$ via the parameter
$\zeta _m$ according to Eq. (\ref{1.22}). The interaction
coefficients $V_c(m)$ appear also directly in $W_2$. In order to
get explicit results for the matrix elements, one must be able to
perform the summations in $\tilde{W}_1$ and $\tilde{W}_2$. This
requires models for the $m$-dependence of the above interaction
coefficients.

Defining $\tilde{V}_c \equiv V_c/\hbar \omega _\ell$, an explicit
form for $\alpha _m$ is

 \begin{equation}\label{4.1}
 \alpha _m = - \frac{\tilde{V}_c (m)}{2
 \sqrt{[1-\tilde{V}_c(m)]^2}}
 \equiv \frac{1-K_m^2}{4 K_m}.
 \end{equation}

Similarly, the central coupling constants $K_m$ determine $\gamma
_m$ and the renormalized oscillator energies $\epsilon _m$
according to

\begin{equation}\label{4.3}
 \gamma _m =  \frac{(1-K_m)^2}{4 K_m},\quad \epsilon _m=\hbar\omega _\ell\,
 \frac{2K_m}{K_m^2+1}.
\end{equation}

Following the procedure in the Luttinger model, we adopt
exponential decays, thus we make the ansatz:

\begin{equation}\label{4.4}
 \alpha _m =  \alpha _0 \exp(- r_\alpha m/2 ),
\end{equation}
 and

\begin{equation}\label{4.5}
 \gamma _m = \gamma _0 \exp (- r_\gamma m).
\end{equation}

 Since the signs of $\tilde{V}_c(m)$ do not depend on $m$, $\alpha _m$
 is related to $\gamma _m$ by

 \begin{equation}\label{4.6}
 \alpha _m = -\mbox{sgn}(\tilde{V}_c)\,\sqrt{\gamma _m (1 + \gamma _m)},
 \end{equation}
 and one finds

\begin{eqnarray}\label{4.7}
\alpha _ m = -\mbox{sgn} (\tilde{V}_c) \,\exp(- r_\gamma m/2)
 \sqrt{\gamma _ 0 [1 + \gamma _ 0 \,\exp(-r_ \gamma m)]}.
 \end{eqnarray}

Assuming that $r_\gamma $ is very small, $0 < r_\gamma \ll 1 $,
and that the relevant values of $m$ obey $ m < 1/r_\gamma $, Eq.
(\ref{4.7}) leads to

 \begin{equation}\label{4.8}
 \alpha _ m \approx \exp(- r_\gamma m/2) \,\alpha _ 0,
 \end{equation}
 with

 \begin{equation}\label{4.9}
 \alpha _0 = - \mbox{sgn} (\tilde{V} _c) \, \sqrt{ \gamma _ 0 (1 + \gamma _ 0 )}.
 \end{equation}
Thus, we can set

 \begin{equation}\label{4.10}
 r_\alpha = r _\gamma \equiv r.
 \end{equation}

 Another set of coupling parameters involves directly the backscattering coefficients
 $\tilde{V}_c(m)$ via $\xi _{2m}$ in $\tilde{W}_2 $.
 Using Eqs. (\ref{1.25}) and (\ref{1.27}), Eq. (\ref{1.35}) becomes

 \begin{equation}\label{4.11}
 \xi _{2m}=- \frac{\tilde{V}_c (m)}{4 m [1 -\tilde{V}_c (2m)]}.
 \end{equation}

We take advantage of the slow decay of interaction coefficients
with $m$ and write

\begin{equation}\label{4.12}
\xi _{2m} =- \frac{1}{4m} \tilde{V}_{ceff} \,\exp(-r_c m),
\end{equation}
with another decay constant $r_c \ll 1$ and with

\begin{equation}\label{4.13}
 \tilde{V} _{ceff} \approx \frac{\tilde{V}_c (1)}{1
 -\tilde{V}_c (1)}\equiv \frac{1}{2}\,(K^2-1).
 \end{equation}
This gives a useful result for the backscattering function
(\ref{1.34})

\begin{equation}\label{4.14}
C(u)=-\frac{\tilde{V} _{ceff}}{4} \, \sum _{m=1}^\infty
\frac{1}{m}\,\exp[(-m(2iu+r_c)] =\frac{\tilde{V}
_{ceff}}{4}\,\ln[1-\exp(-r_c-2iu)].
\end{equation}
It is noted that consistency requires $|\tilde{V}_c |\le 1$ and $K
\ge 0$.

Comparing $\alpha(m)$ in Eq. (\ref{4.1}) with Eqs. (\ref{4.11})
and (\ref{4.12}), leads to

\begin{equation}\label{4.19}
 \alpha _m = - \tilde{V}_{ceff}\, \exp(-r_c m) \left \{\frac{ 1-\tilde{V}_c (2m)}
 {2 \sqrt{1- \tilde{V} _c^2 (m)}} \right \}.
\end{equation}
Suppressing the weak $m$-dependence in the curly brackets, i.e.,

 \begin{equation}\label{4.20}
\tilde{V}_c (m)\rightarrow \tilde{V}_c,
\end{equation}
 allows the identifications

 \begin{equation}\label{4.21}
 r_c = \frac{r_\alpha}{2}=\frac{r}{2}.
 \end{equation}

 Furthermore, using

 \begin{equation}\label{4.22}
 \alpha _0 \equiv - \frac{\tilde{V}_c}{2 \sqrt{1-\tilde{V}^2_c}},
 \end{equation}
 gives

 \begin{eqnarray}\label{4.23}
 \tilde{V} _{ceff} =-2 \,\sqrt{\frac{1+\tilde{V}_c}{1- \tilde{V}_c}}\,\,\alpha _0.
 \end{eqnarray}

The main coupling constant

 \begin{eqnarray}\label{4.24}
 K=\sqrt{\frac{1+\tilde{V}_c}{1-\tilde{V}_c}}
 \end{eqnarray}
 determines all relevant couplings and the renormalized energy according to

 \begin{eqnarray}\label{4.25}
\tilde{V}_c =\frac{K^2-1}{K^2+1},\quad \alpha
_0=-\frac{K^2-1}{4K},\quad \gamma _0 = \frac{(K-1)^2}{4K},\quad
\epsilon=\hbar\omega _\ell\,\frac{2K}{K^2+1}.
\end{eqnarray}

Only the two parameters $K$ and $r=r_\alpha =r_\gamma =2 r_c$
remain. $K$ is determined by the physical backscattering strength
$\tilde{V}_c(1) \equiv \tilde{V}_c$ according to Eq. (\ref{4.24}).
In \cite{GaW02}, it was argued that $r = 1/\sqrt{N}$ is a
reasonable choice for the decay constant. Finally, it is noted
that interactions always decrease the renormalized excitation
energy $\epsilon$ below the non-interacting value $\hbar \omega
_\ell$.


\begin{thebibliography}{111111}
\bibitem{WWW}
The "proof" indicated in \cite{WW01} to show that $S(N,\alpha _1)=N$ is flawed. It used
an interchange of limiting procedures which is wrong, especially at strong coupling.
\end{thebibliography}
\end{document}